\documentclass[twoside]{dis04}
\def\runauthor{Strikman and Weiss}
\def\shorttitle{Gluonic transverse size}
\usepackage{bm} 
\usepackage{graphicx}
\begin{document}
\title{The nucleon's gluonic transverse size: \\
From exclusive $J/\psi$ photoproduction to
high--energy $pp$ collisions}
\author{M. Strikman}
\address{Pennsylvania State University, University Park, PA 16802, U.S.A.}
\author{C. Weiss}
\address{Theory Group, Jefferson Lab, Newport News, VA 23606, U.S.A.}
\maketitle
\abstracts{We summarize what is known about the transverse spatial
distribution of gluons in the nucleon and its $x$--dependence from 
exclusive $J/\psi$ photo/electroproduction 
in $ep$ fixed--target and collider experiments (HERA H1 and ZEUS).
This information can be used to predict the impact parameter 
dependence of the cross section for certain hard QCD processes
(dijet production) in $\bar pp$ and $pp$ collisions at the Tevatron 
and LHC.}
The parton picture of the nucleon deals not only with the 
distribution of partons with respect to longitudinal momentum, 
but also with their spatial distribution in the transverse plane. 
The latter aspect, which was investigated by Gribov in a general 
context \cite{Gribov:jg}, is receiving new attention in connection 
with the recent interest in generalized parton distributions (GPD's). 
The spatial distribution of partons 
in the transverse plane can be resolved in hard exclusive processes, 
in which the nucleon is observed in the
final state, with an invariant momentum transfer $t \ll \mu^2$
($\mu^2$ is the hard scale characterizing the process).
Examples include the photoproduction of heavy quarkonia ($J/\psi$),
or the hard electroproduction of light mesons or real photons 
(deeply virtual Compton scattering). At the same time, the transverse 
spatial distribution of partons plays an important role in the description 
of hard QCD processes in hadron--hadron collisions at high 
energies \cite{Frankfurt:2003td}. This concept thus provides a means 
by which information gained in $ep$ scattering experiments 
can be used to make predictions for $\bar pp$ and $pp$ collisions 
at the Tevatron and the LHC.

The generalized parton distributions parametrize the non-diagonal matrix
elements of the twist--2 QCD quark and gluon operators between nucleon
states. In the simplest case, when the momentum transfer between the
nucleon states has only a transverse component, $\bm{\Delta}_\perp$,
the generalized gluon distribution, $H_g (x, t)$, with 
$t \equiv -\bm{\Delta}_\perp^2$, can be interpreted as the elastic 
nucleon form factor for gluons with longitudinal momentum fraction $x$,
with $H_g (x, t = 0) = g(x)$ the usual gluon density.
One can represent this form factor as the Fourier transform 
of a function of a transverse coordinate, $\bm{\rho}$,
\begin{equation}
H_g (x, -\bm{\Delta}_\perp^2 ) \;\; = \;\; \int d^2 \rho \; 
e^{i (\bm{\Delta}_\perp \bm{\rho})} \; g(x, \rho)
\hspace{4em}
(\rho \equiv |\bm{\rho}|).
\label{impact_def}
\end{equation}
This function describes the spatial distribution
of gluons with longitudinal momentum fraction $x$ in the transverse 
plane \cite{Burkardt:2002hr}. Its integral over the 
transverse coordinate gives back the total gluon density,
\begin{equation}
\int d^2\rho \; g(x, \rho) \;\; = \;\; g(x) .
\end{equation}
In particular, the average ``gluonic transverse size'' of the nucleon
is related to the slope of the generalized gluon distribution at $t = 0$,
\begin{equation}
\langle \rho^2 \rangle
\;\; \equiv \;\; \frac{\int d^2 \rho \; g(x, \rho) \; \rho^2}
{\int d^2 \rho \; g(x, \rho)} 
\;\; = \;\; 4 \; \frac{\partial}{\partial t}
\left[ \frac{H_g (x, t)}{H_g (x, 0)} \right]_{t = 0} . 
\end{equation}
This quantity allows for an intuitive physical interpretation,
and can be compared with other measures of the 2--dimensional size 
of the nucleon, for example, with 2/3 times the 3--dimensional electric 
or axial charge radius squared of the nucleon.

The gluonic transverse size of the nucleon is expected to grow with 
decreasing longitudinal momentum fraction, $x$. Different 
physical mechanisms are responsible for this growth in different 
regions of $x$. At $x \rightarrow 1$, $\langle \rho^2 \rangle$ vanishes 
because the $t$--dependence of the generalized gluon distribution
disappears if one parton carries the entire longitudinal momentum of 
the nucleon (Feynman mechanism).
When $x$ is decreased below the valence region, a distinctive increase 
of $\langle \rho^2 \rangle$ is caused by pion cloud contributions
to the gluon density, which set in for $x < M_\pi / M_N$
\cite{Strikman:2003gz,Frankfurt:2002ka}.
Finally, when $x$ is decreased further, the transverse size grows
due to the random walk character of successive emissions 
in the partonic ladder (Gribov diffusion) \cite{Gribov:jg}.

The transverse spatial distribution of gluons also changes with 
the scale, $\mu^2$, as a result of DGLAP evolution. For $\mu^2$
sufficiently large compared to the transverse spatial resolution, 
$\mu^2 \gg 1/(\Delta\rho )^2$, the parton decays happen
locally in transverse position. Looking at the
$\rho$--distribution at some fixed $x$, one finds that that it shrinks
with increasing $\mu^2$. The reason is that, as $\mu^2$ increases,
the distribution becomes sensitive to the input distribution 
(at the initial scale) at higher values of $x$, 
where it is concentrated at smaller transverse 
distances \cite{Frankfurt:2003td}.

Most of the experimental information about the transverse 
spatial distribution of gluons in the nucleon comes from 
exclusive $J/\psi$ photoproduction, $\gamma + N \rightarrow 
J/\psi + N$. At the leading twist level, the amplitude for
this process is proportional to the generalized gluon 
distribution in the nucleon, with 
$\mu^2 \sim (\mbox{size of $\bar c c$})^{-2} \approx 3 \, {\rm GeV}^2$, 
and $x \sim (\mbox{mass of $\bar c c$})^2 / W^2$ for sufficiently 
large $W \equiv \sqrt{s}$ (for smaller energies the ``skewedness'' of the 
gluon distribution, $x_1 \neq x_2$, becomes important) 
\cite{Frankfurt:1997fj}. In this approximation the $t$--dependence 
of the generalized gluon distribution can directly be inferred from 
the measured $t$--dependence of the differential cross section.

\begin{figure}[t]
\includegraphics[width=10cm,height=7cm]{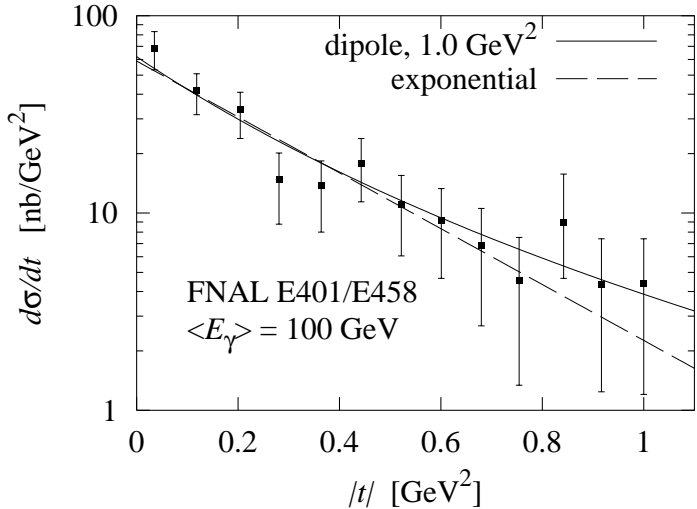}
\caption[*]{The $t$--dependence of the differential cross section
for exclusive $J/\psi$ photoproduction measured in the FNAL E401 / E458 
experiment \cite{Binkley:1981kv}. Dashed line: Exponential fit
$e^{3.26 t/ {\rm GeV}^2}$. Solid line: Dipole 
parametrization $(1 - t/1.0 \, {\rm GeV}^2)^{-4}$.}
\label{fig_fnal458}
\end{figure}

Exclusive $J/\psi$ photoproduction has been studied 
in a number of fixed--target experiments: 
Cornell at $E_\gamma = 11.8\; {\rm GeV}$ \cite{Gittelman:1975ix}, 
SLAC at $E_\gamma = 19\; {\rm GeV}$ \cite{Camerini:1975cy}, 
CERN NA14 at $\langle E_\gamma \rangle = 90\; {\rm GeV}$
\cite{Barate:1986fq}, and FNAL E401/E458 at 
$\langle E_\gamma \rangle = 100\; {\rm GeV}$ 
\cite{Binkley:1981kv} (in this experiment also the recoiling proton 
was detected). Fig.~\ref{fig_fnal458} shows the $t$--dependence
of the differential cross section measured in the FNAL E401/E458 
experiment. An exponential fit $d\sigma / dt \propto e^{Bt}$ gives
$B = 3.26 \pm 1.30 \; {\rm GeV}^{-2}$. In order to extract the gluonic 
transverse size of the nucleon one should correct for the 
finite size of the produced $\bar c c$ system, which leads to an 
additional ``smearing'' in the transverse plane and accounts
for approximately $0.3 \; {\rm GeV}^{-2}$ of the 
observed $B$ value \cite{Frankfurt:1997fj}. 
Subtracting this contribution we obtain the estimate
\begin{equation}
\langle \rho^2 \rangle \;\; \approx \;\; 0.24 \, {\rm fm}^2 
\hspace{4em} (x \sim 10^{-1}).
\label{rho2_exponential}
\end{equation}

One expects the $t$--dependence of the generalized gluon distribution 
at $x \ge 10^{-1}$ to be similar to the nucleon's axial form 
factor \cite{Frankfurt:2002ka}. The reason is that, like the axial form 
factor, the generalized gluon distribution for $x > M_\pi / M_N$ 
does not receive contributions from the nucleon's pion cloud
\cite{Strikman:2003gz,Frankfurt:2002ka}. The $t$--dependence should 
thus be well described by the dipole parametrization (see 
Ref.\cite{Bernard:2001rs} for a review)
\begin{equation}
H_g (x, t) \;\; \propto \;\; (1 - t/m_g^2)^{-2},
\hspace{4em} m_g^2 \;\; = \;\; 1.1 \, {\rm GeV}^2 
\hspace{4em} (x \sim 10^{-1}).
\label{dipole}
\end{equation}
This corresponds to a gluonic transverse size of the nucleon of
\begin{equation}
\langle \rho^2 \rangle \;\; = \;\; 8/m_g^2 
\;\; \approx \;\; 0.28 \, {\rm fm}^2 
\hspace{4em} (x \sim 10^{-1}),
\label{rho2_dipole}
\end{equation}
consistent with the estimate (\ref{rho2_exponential}).
The dipole parametrization (\ref{dipole}) implies a 
$t$--dependence of the differential cross section as 
$(1 - t/1.0 \, {\rm GeV}^2)^{-4}$, where
the 10\% decrease in the mass parameter accounts for the 
finite size of the $\bar c c$ system (see above). This form 
describes well the E401/E458 data, see 
Fig.~\ref{fig_fnal458}. It also describes the 
$t$--dependence of the data at much lower 
energies \cite{Gittelman:1975ix,Camerini:1975cy}; 
see Ref.~\cite{Frankfurt:2002ka} for details.
In particular, in this way the observed decrease of the $B$ 
parameter in exponential fits to the low--energy data can be 
explained as the result of sampling the dipole form factor 
at larger $|t| > |t_{\rm min}|$, where its logarithmic slope 
becomes smaller.

$J/\psi$ photoproduction near threshold ($E_\gamma = 8.2\; {\rm GeV}$) 
will be investigated at Jefferson Lab Hall A with the 11 GeV 
electron beam \cite{Chudakov:2001eu}. An interesting question
is whether the two--gluon exchange mechanism will still be applicable
in this region, or whether a coherent multi--parton reaction 
mechanism will take over \cite{Brodsky:2000zc}. This issue will be 
crucial also for interpreting the data on sub--threshold $J/\psi$ 
photoproduction off nuclei expected from the E-03-008 experiment at 
Jefferson Lab Hall C.

$J/\psi$ photoproduction has also been studied
extensively by the H1 and ZEUS experiments at the HERA collider.
The effective $x$ values in the generalized gluon distribution
here are $x \sim 10^{-2} - 10^{-3}$. The $t$--dependence 
of the measured differential cross section in both experiments
is well described by the exponential form, $e^{Bt}$. 
H1 quotes a value of $B = 4.73 \pm 0.25^{+0.30}_{-0.39} \; {\rm GeV}^{-2}$ 
for data averaged over the range $40 < W < 150\; {\rm GeV}$ 
\cite{Adloff:2000vm}.
The recent ZEUS analysis of $J/\psi$ electroproduction
data reports a value of $4.72 \pm 0.15 \pm 0.12 \; {\rm GeV}^{-2}$ 
for the combined data in the range $2 < Q^2 < 100 \; {\rm GeV}^2$,
with no noticeable $Q^2$ dependence \cite{:2004mw}. 
The previous ZEUS photoproduction data suggested a somewhat smaller value,
$B = 4.15 \pm 0.05^{+0.30}_{-0.18} \; {\rm GeV}^{-2}$ at 
$W = 90\; {\rm GeV}$ \cite{Chekanov:2002xi}. 
Correcting again for the finite size 
of the $\bar c c$ system, we estimate the gluonic transverse size 
of the nucleon as
\begin{equation}
\langle \rho^2 \rangle \;\; \approx \;\; 0.35 \, {\rm fm}^2 
\hspace{4em} (x \sim 10^{-2} - 10^{-3}).
\label{rho2_HERA}
\end{equation}
This is about 30\% larger than the estimate (\ref{rho2_exponential})
at $x \sim 10^{-1}$.

\begin{figure}[t]
\includegraphics[width=10cm,height=7cm]{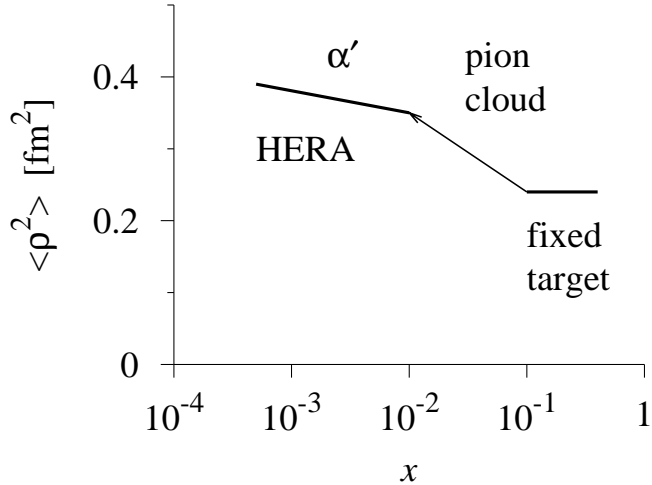}
\caption[*]{Schematic illustration of the $x$--dependence
of the gluonic transverse size of the nucleon, $\langle \rho^2\rangle$.
The increase between $x \sim 10^{-1}$ and $x \sim 10^{-2}$
can be attributed to the contribution of the nucleon's pion cloud
to the gluon density.}
\label{fig_rho2}
\end{figure}

The increase in the nucleon's gluonic transverse size between 
fixed--target and collider energies can be explained as the 
result of the contribution of the nucleon's pion cloud to the gluon 
distribution $g(x, \rho )$ at distances $\rho \sim 1/M_\pi$, 
which is suppressed for $x > M_\pi / M_N \sim 10^{-1}$ but becomes 
noticeable around $x \sim 10^{-2}$. It was estimated in 
Ref.~\cite{Strikman:2003gz} that this results in an increase 
of the gluonic transverse size of the nucleon of
\begin{equation}
\langle \rho^2 \rangle_{\mbox{\scriptsize pion cloud}} 
\;\; \approx 0.06 \; {\rm fm}^2 ,
\end{equation}
which is roughly consistent with the difference between 
the estimates (\ref{rho2_HERA}) and (\ref{rho2_exponential}), 
(\ref{rho2_dipole}). The H1 and ZEUS 
experiments have measured also the change of the logarithmic $t$--slope 
of the cross section with the CM energy, $\alpha'$, over the $W$ range
covered in these experiments. In our context this parameter can be 
interpreted as the change of the nucleon's gluonic transverse size 
with $\ln (1/x)$,
\begin{equation}
\frac{1}{4} \; \frac{\partial \; \langle \rho^2 \rangle}
{\partial\ln (1/x)}
\;\; = \;\; \alpha' .
\label{alphap}
\end{equation}
The H1 photoproduction data estimates a value of 
$\alpha' = 0.08 \pm 0.17\; {\rm GeV}^{-2}$ \cite{Adloff:2000vm}; 
the new ZEUS analysis of electroproduction data quotes
$0.07 \pm 0.05 (\mbox{stat}) { }^{+0.03}_{-0.04}
(\mbox{syst}) \; {\rm GeV}^{-2}$ \cite{:2004mw}. Thus, the variation
of $\langle\rho^2\rangle$ in the range $x \sim 10^{-3} - 10^{-2}$
is rather small, and the extrapolation to the region $x \sim 10^{-1}$ 
does not match with the value extracted from the fixed target data.
This indicates that the change in $\langle\rho^2\rangle$ indeed 
happens rather suddenly between $x \sim 10^{-1}$ and $x \sim 10^{-2}$, 
as implied by the pion cloud mechanism. Fig.~\ref{fig_rho2} 
schematically illustrates this scenario.

\begin{figure}[t]
\includegraphics[width=10cm,height=7cm]{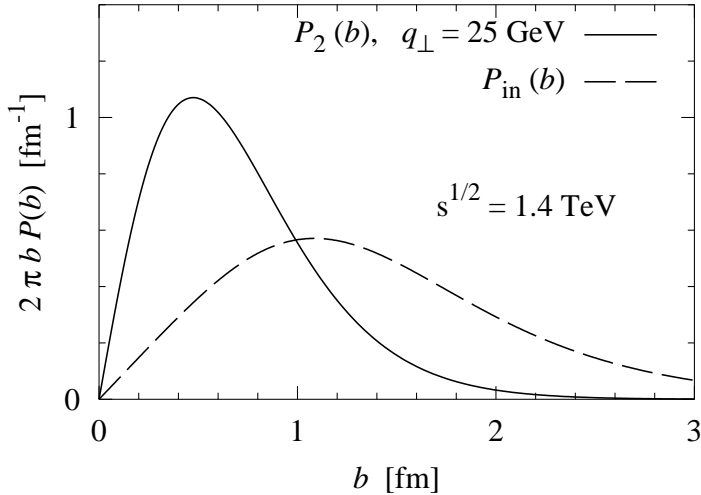}
\caption[*]{Solid line: Probability for producing a hard dijet
($q_\perp = 25\; {\rm GeV}$) in $pp$ collisions at LHC,
$P_2 (b)$ of Eq.~(\ref{P_2}), as a function of the $pp$ 
impact parameter, $b$. Dashed line: Probability distribution
for generic inelastic events, $P_{\rm in}(b)$. Shown are the 
radial distributions, $2\pi b \, P(b)$.}
\label{fig_pb}
\end{figure}
The information about the transverse spatial distribution of gluons 
gained from the study of exclusive processes in $ep$ scattering
can be used to make predictions for certain characteristics 
of $\bar pp$ and $pp$ collisions at high energies (Tevatron, LHC). 
In particular, one can predict the probability for the 
production of a hard dijet at zero rapidity 
by a gluon--gluon collision, depending on the 
impact parameter of the underlying $pp$ collision, 
$b$ \cite{Frankfurt:2003td}.
\begin{equation}
P_2 (b) \;\; = \;\; \int d^2\rho_1 \int d^2\rho_2 \; 
\delta^{(2)} (\bm{b} - \bm{\rho}_1 + \bm{\rho}_2 )
\; \frac{g (x, \rho_1 )}{g(x)} \; 
\frac{g (x, \rho_2 )}{g(x)} ,
\label{P_2}
\end{equation}
where $x = 2 q_\perp / \sqrt{s}$ is the momentum fraction of the
colliding gluons, with $q_\perp$ the transverse momentum of the dijet.
Fig.~\ref{fig_pb} shows $P_2(b)$ for dijets with 
$q_\perp = 25\; {\rm GeV}$ in $pp$ collisions at the LHC 
($\sqrt{s} = 14 \; {\rm TeV}$), as obtained from the parametrization 
of $g(x, \rho)$ of Ref.~\cite{Frankfurt:2003td}, which is
based on the dipole parametrization (\ref{dipole}) and incorporates 
the information about the $x$--dependence of $\langle\rho^2\rangle$
shown in Fig.~\ref{fig_rho2}. This $b$--distribution is much narrower 
than the corresponding distribution for generic inelastic events 
(i.e., with no condition on hard processes), 
$P_{\rm in}(b)$, which can be inferred from the impact parameter 
representation of the amplitude for $pp$ elastic scattering. 
The reason is that the cross section for generic inelastic events 
is dominated by collisions of soft partons, whose transverse spatial
distribution is much wider than that of the hard partons required
to make the dijet. Conversely, this means that requiring the
presence of a hard dijet (e.g., by way of a trigger) amounts to a
``filter'' for central $pp$ collisions at high 
energies \cite{Frankfurt:2003td}.
This possibility is of considerable practical importance.
In particular, it allows systematic studies of the approach to the
unitarity (``black body'') limit in central $pp$ collisions, which 
would manifest itself in certain modifications of particle production
at forward/backward rapidities. Such studies could be performed
with the CMS/TOTEM detectors at LHC. Finally, we note that the
transverse spatial distribution of large--$x$ partons plays an important 
role also in the diffractive production of Higgs bosons
at LHC \cite{Higgs}.

The ideas presented here were developed in collaboration with 
L.~Frankfurt. We thank E.~Chudakov for interesting discussions. 
This work was supported by the US Department of Energy contract
DE-AC05-84ER40150, under which the Southeastern Universities Research 
Association (SURA) operates the Thomas Jefferson National Accelerator Facility.
The work of C.~W.\ was partly supported by Deutsche Forschungsgemeinschaft 
(Heisenberg Fellowship).

\begin{thebibliography}{0}
%
%
\bibitem{Gribov:jg}
V.~N.~Gribov, arXiv:hep-ph/0006158 .
%
%
\bibitem{Frankfurt:2003td}
L.~Frankfurt, M.~Strikman and C.~Weiss,
Phys.\ Rev.\ D {\bf 69}, 114010 (2004).
%
%
\bibitem{Burkardt:2002hr} 
M.~Burkardt,
Int.\ J.\ Mod.\ Phys.\ A {\bf 18}, 173 (2003); Phys.\ Rev.\ D {\bf 66}, 
114005 (2002). P.~V.~Pobylitsa, Phys.\ Rev.\ D {\bf 66}, 094002 (2002).
%
%
\bibitem{Strikman:2003gz}
M.~Strikman and C.~Weiss,
Phys.\ Rev.\ D {\bf 69}, 054012 (2004)
%
%
\bibitem{Frankfurt:2002ka}
L.~Frankfurt and M.~Strikman,
Phys.\ Rev.\ D {\bf 66}, 031502 (2002).
%
%
\bibitem{Frankfurt:1997fj}
L.~Frankfurt {\it et al.}, Phys.\ Rev.\ D {\bf 57}, 512 (1998).
L.~Frankfurt {\it et al.}, JHEP {\bf 9902}, 002 (1999); 
JHEP {\bf 0103}, 045 (2001).
%
%
\bibitem{Gittelman:1975ix}
B.~Gittelman {\it et al.}, 
Phys.\ Rev.\ Lett.\  {\bf 35}, 1616 (1975).
%
%
\bibitem{Camerini:1975cy}
U.~Camerini {\it et al.},
Phys.\ Rev.\ Lett.\  {\bf 35}, 483 (1975).
%
%
\bibitem{Barate:1986fq}
R.~Barate {\it et al.}  [NA14 Collaboration],
Z.\ Phys.\ C {\bf 33}, 505 (1987).
%
%
\bibitem{Binkley:1981kv}
M.~Binkley {\it et al.},
Phys.\ Rev.\ Lett.\  {\bf 48}, 73 (1982).
%
%
\bibitem{Bernard:2001rs}
V.~Bernard, L.~Elouadrhiri and U.~G.~Meissner,
J.\ Phys.\ G {\bf 28}, R1 (2002).
%
%
\bibitem{Chudakov:2001eu}
E.~Chudakov {\it et al.}, JLAB-TN-01-007
%
%
\bibitem{Brodsky:2000zc}
S.~J.~Brodsky {\it et al.}, Phys.\ Lett.\ B {\bf 498}, 23 (2001).
%
%
\bibitem{Adloff:2000vm}
C.~Adloff {\it et al.}  [H1 Collaboration],
Phys.\ Lett.\ B {\bf 483}, 23 (2000).
%
%
\bibitem{:2004mw} [ZEUS Collaboration],
arXiv:hep-ex/0404008 .
%
%
\bibitem{Chekanov:2002xi}
S.~Chekanov {\it et al.}  [ZEUS Collaboration],
Eur.\ Phys.\ J.\ C {\bf 24}, 345 (2002).
%
%
\bibitem{Higgs} L.~Frankfurt, M.~Strikman and C.~Weiss, in preparation.
%
%
\end{thebibliography}
\end{document}